\documentstyle[12pt]{article}
\def\e {\epsilon}
\def\be{\begin{equation}}
\def\ee{\end{equation}}
\def\ba{\begin{array}}
\def\ea{\end{array}}
\def\EQ{\begin{equation}}
\def\EN{\end{equation}}
\def\bea{\begin{eqnarray}}
\def\eea{\end{eqnarray}}
\def\to{\rightarrow}

\def\sb{\hspace{0.2in}}

\def\O{{\cal O}}

\def\K{{\cal K}}
\def\M{{\cal M}}

\begin{document}
\oddsidemargin 5mm
\renewcommand{\thefootnote}{\fnsymbol{footnote}}
\newpage
\begin{flushright}
DAMTP-HEP-95/26
\end{flushright}
\vspace{0.2cm}
\begin{center}
{\large {\bf Local Operators in Massive Quantum Field Theories}
}\\
\vspace{0.3cm}
{\bf
Anni Koubek\footnote{Work supported by PPARC grant no. GR/J20661} }\\
\vspace{0.2cm}
{\em Department of Applied Mathematics and Theoretical Physics, \\
CB3 9EW Cambridge, UK }
\end{center}

\subsection*{\large\bf 1 \phantom{11} Introduction}

A fundamental problem in quantum theory is to establish a
connection between its local description (quantum field theory)
and measurable
quantities (particle masses, scattering amplitudes). In elementary
particle physics one relies mostly on approximation techniques due to
the non-integrable structure of the interactions. In order to gain a
deeper understanding of quantum theory in general these issues are
examined for integrable systems, where one hopes to gain exact
relations between the two descriptions. Though many integrable
theories in four dimensions are known (for a review see e.g. \cite{fourd})
much more knowledge has been obtained so far for two dimensional
theories, and it is thus this class of models where my investigations
focus.

In the particle picture the interaction is encoded into the scattering
matrix. The asymptotic states are described by a linear superposition
of free one-particle states $\vert Z_\e (\beta)\rangle$, which are
characterised by the particle species $\e$ and their momentum,
parametrised as $p^{(0)} =m \cosh \beta$, $p^{(1)} =m \sinh \beta$
($m$ denotes the mass and $\beta$ the rapidity).  They are related
through the $S$-matrix as \be \vert Z_{\e_1}(\beta_1) \dots
Z_{\e_n}(\beta_n) \rangle_{in} =
S_{\e_1\dots\e_n}^{\e'_1\dots\e'_m}(\beta_1,\dots,\beta_n \vert
\beta_1'\dots\beta'_m ) \, \vert Z_{\e'_1}(\beta'_1) \dots
Z_{\e'_m}(\beta'_m) \rangle_{out} \sb .\ee On the other hand the local
description of a theory consists of the space of local operators
${\cal A} = \{ \O_i \} $ and the set of multi-point correlation
functions of them, $$ \langle 0\vert \O_1(x_1 ) \dots \O_n(x_n ) \vert
0 \rangle \sb .$$ The two description are linked through the
Lehmann-Symanzik-Zimmermann reduction, since the particles can be
obtained as asymptotic limits of the local fields.  Another connection
is given through the form factors. Consider an arbitrary two-point
correlation function

\[
G_{ij}(x)\,=\,<\O_i(x)\,\O_j(y)> \sb ,
\]
 of hermitian operators. Inserting the Identity between the two
 operators and expanding it into the base of asymptotic states,
it can be expressed as an infinite series
over multi-particle intermediate states,

\be
\langle \O_i(x)\,\O_j(y)\rangle\,=\label{correlation}\ee
$$ \sum_{n=0}^{\infty}
\int \frac{d\beta_1\ldots d\beta_n}{ (2\pi)^n}
<0|\O_i(x)|Z_{\e_1}(\beta_1),\ldots,Z_{\e_n}(\beta_n)>_{\rm in}{}^{\rm
in} <Z_{\e_1}(\beta_1),\ldots,Z_{\e_n}(\beta_n)|\O_j(y)|0>  $$

The matrix elements
$$<0|\O_i(0)|Z_{\e_1}(\beta_1),\ldots,Z_{\e_n}(\beta_n)> =
F_{\e_1\dots\e_n}(\beta_1,\dots,\beta_n)$$
are called form factors and in the following I will try to explain how
they can be used in order to establish a link between the local
description and the particle picture of a theory.

\subsection*{\large\bf 2 \phantom{11}
Form Factors and the Space of Local Operators}

If one considers two dimensional integrable theories many simplifying
properties occur which allow to calculate many dynamical quantities
exactly. The most remarkable fact is the factorisation of the
$S$-matrix, which determines a general scattering process as a product
of two-particle scattering amplitudes. Further these two-particle
$S$-matrices are pure phase-shifts, that is the incoming and outgoing
momenta are the same. This simplification allows to calculate the
$S$-matrix exactly (see {\em e.g.} P. Kulish's lectures in these proceedings).

Also the form-factors can be determined exactly for integrable two
dimensional systems. They obey a set of constraint equations,
originating from fundamental principles of quantum theory, such as
unitarity, analyticity, relativistic covariance and locality
\cite{Karowski,nankai}.
The important fact is that the  $S$-matrix is the only dynamical
information needed.
In the following I will discuss just two examples of form-factor
equations, in order to determine their overall structure, and also to
explain the solution techniques.

Since the theories here considered are defined in only one space dimension, a
scattering process can be viewed as to interchange two particles on the
real line,
\be
Z_{\epsilon_1}(\beta_1) Z_{\epsilon_2}(\beta_2) =
S_{\epsilon_1\epsilon_2}(\beta_1-\beta_2)
Z_{\epsilon_2}(\beta_2) Z_{\epsilon_1}(\beta_1)
\sb .\ee
This exchange property will lead to a constraint equation for the
form-factors
\bea
& & F^\O_{\e_1\dots\e_i\e_{i+1}\dots\e_n}
(\beta_1,\dots,\beta_i,\beta_{i+1},\dots\beta_n) =\nonumber\\ &
&S_{\e_i\e_{i+1}}(\beta_i-\beta_{i+1})\,
F^\O_{\e_1\dots\e_{i+1}\e_{i}\dots\e_n}
(\beta_1,\dots,\beta_{i+1},\beta_{i},\dots\beta_n) \sb .
\label{wat1}\eea

Another constraint equation derives from the bound state structure of
the theory under consideration. If particles $Z_i$, $Z_j$ form a bound
state $Z_k$, the corresponding
two-particle scattering amplitude
 exhibits a pole at $\beta = iu_{ij}^{k}$ with the residue
\be
 -i \lim_{\beta \to iu_{ij}^{k}}(\beta-iu_{ij}^{k})
S_{ij}(\beta) =(\Gamma_{ij}^{k})^2 \sb ;
\label{residueS}\ee
$\Gamma_{ij}^{k}$ is the three--particle on--shell vertex.
Corresponding to this bound state the form--factor exhibits a pole
 with the
residue $$ -i \lim_{\beta'\to \beta}(\beta'-\beta)
F^\O_{\e_1\dots ij\dots\e_n} (\beta_1,\dots,\beta'+i(\pi-
u_{ik}^{j}),\beta-i(\pi- u_{jk}^
{i}),\dots,\beta_{n}) = $$
\be =\Gamma_{ij}^{k} F^\O_{\e_1\dots k\dots\e_n}
 (\beta_1,\dots,\beta,\dots,\beta_{n} )
\sb . \label{bounds}\ee

As mentioned before, (\ref{wat1}) and (\ref{bounds}) are only two
examples of form-factor equations. Nevertheless they are two exponents
of the two categories of the constraint equations:
\begin{enumerate}
\item Equations with fixed n ( e.g. (\ref{wat1})): they involve form
  factors with the same number of particles on both sides of the
  equation
\item Recursive equations  ( e.g.(\ref{bounds})): They link form
  factors with different particle numbers with each other - in the
  example above $n+1$ particle form factors to $n$ particle form
  factors.
\end{enumerate}

For theories with scalar particles there exists a well established
solution method \cite{Karowski}. It consists of the ansatz
\be
F_{\e_1\dots\e_n}(\beta_1,\dots,\beta_n)
=Q_{\e_1\dots\e_n}(e^{\beta_1},\dots,e^{\beta_n})
\prod_{i<j}^n F_{\e_i\e_j} (\beta_i -\beta_j)\sb . \ee
The two-particle form factors $F_{\e_i\e_j}$
 can be calculated easily from the form
factor equations. The product term satisfies all equations of the
first type (with fixed particle number) and also is designed in order
to have the correct pole structure of an $n$-particle form factor.

Through this parametrisation the form-factor equations are reduced to
recursive relations for the functions
$Q_{\e_1\dots\e_n}(e^{\beta_1},\dots,e^{\beta_n})$. Further properties
of these functions can be extracted from the form factor equations:
they are homogeneous polynomials, symmetric in repeated indices with a total
degree in its arguments fixed by relativistic covariance and the
partial degree determined from the recursion relations.

This information is sufficient for simple models to obtain explicit
expressions for the form factors.  In all cases though it is possible
to determine the space of local operators by just considering these
general properties of the functions $Q$ \cite{mywork}.  Each linear
independent solution of the form factor equations corresponds to an
independent local operator.  Therefore the space of local operators
can be determined by counting the number of independent solutions of
the form factor equations.  This  can be done due to the
 property of the recursion relations, that
the dimension of the solution space at level $n$ is the sum of the
dimension of the solution space at level $n-1$ and of the dimension of
the kernel of the recursion relation at level $n$. Symbolically this
can be written as
$$dim(Q_n) = dim(Q_{n-1}) + dim ( \K_n)\sb .$$

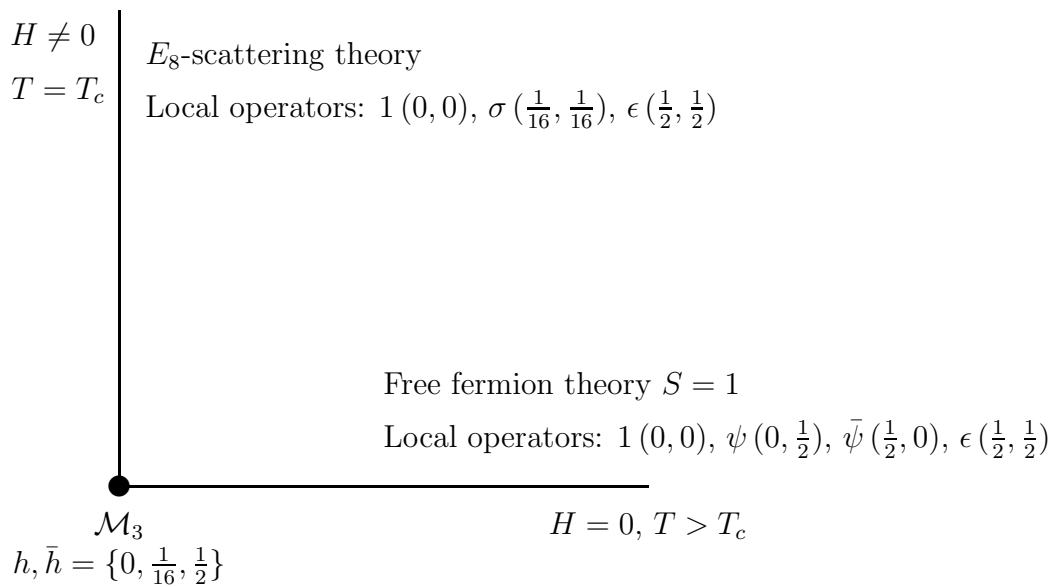
\begin{figure}[hbt]
\begin{center}
\begin{picture}(330,200)(-30,-30)
\thicklines
\put(0,0){\circle*{8}}
\put(0,0){\line(1,0){200}}
\put(0,0){\line(0,1){180}}
\put(0,-15){\makebox(0,0){$\M_{3}$}}
\put(0,-30){\makebox(0,0){$h,\bar{h} =\{ 0,\frac 1{16},\frac 12\}$}}
\put(200,-15){\makebox(0,0){$H=0$, $T>T_c$}}
\put(100,35){Free fermion theory $S=1$}
\put(100,15){Local operators: $1 \, (0,0)$,  $\psi \, (0,
  \frac 12)$,  $\bar{\psi} \, (
  \frac 12,0)$, $\epsilon \, (\frac 12, \frac 12)$}
\put(-25,170){\makebox(0,0){$H \neq 0$}}
\put(-25,150){\makebox(0,0){ $T=T_c$}}
\put(10,160){$E_8$-scattering theory}
\put(10,140){Local operators: $1 \, (0,0)$, $\sigma \, (\frac 1{16},
  \frac 1{16})$, $\epsilon \, (\frac 12 ,\frac 12)$}
\end{picture}
\caption{The critical Ising model and its integrable perturbations}
\label{fig2}
\end{center}
\end{figure}

An interesting application of this counting method is in perturbed
conformal field theories. As an example consider the Ising model. Its
point of second order phase transition is described by a conformal
field theory ( the minimal model $\M_{3}$ ) and therefore the space of
local operators at the critical point is determined by Virasoro
irreducible representations. The critical point admits two relevant
perturbations which are integrable. The perturbation with the
conformal operator with conformal weight $h=\frac 12$ drives the model
into the regime $T>T_c$ and $H=0$. It is described by a free fermion
theory. The other perturbation with the operator $h=\frac 1{16}$
corresponds to the Ising model with $H \neq 0$ and $T=T_c$ and is
described by a scattering theory containing 8 scalar massive particles
\cite{Zam}. Virasoro symmetry is obviously broken by both perturbations
and it is therefore an interesting problem to determine the space of
operators for these theories.

The situation is summarised in figure \ref{fig2}. For both
perturbations it is possible to determine the space of local
operators. It is given in terms of characters of the minimal model
$M_{3}$. This is a quite remarkable fact, since conformal symmetry is
explicitly broken by the perturbation. Further, since the thermal
perturbation is described by a free fermion theory, the local
operators are just the fermions $(\psi, \bar{\psi})$ the identity
operator ($1$) and the energy density $\epsilon$. Also the spin operator
($\sigma$) and the disorder field ($\mu$) can be analysed by the
counting method, but they are semi-local operators with respect to the
asymptotic states. The magnetic perturbation breaks the $Z_2$ symmetry
of the model and only scalar operators appear in this perturbation,
namely the identity ($1$), the energy density ($\epsilon$) and the spin
operator ($\sigma$).

A general feature of the counting method is that the space of
operators is determined by fermionic sum expressions. Such expressions
also appear in the analysis of corner transfer matrices \cite{kedem}
and spinon conformal field theories \cite{spinon}.  It would be
interesting to establish a more direct connection between these
methods and the form factor approach.
Finally note that the above example only constitutes a simple
application of the counting method. It can be generalised to many
other systems, including models with a massless spectrum and/or
bounderies.


\begin{thebibliography}{99}
  \footnotesize \baselineskip=7pt
\bibitem{fourd}  C. Devchand,
  V. Ogievetskii, {\em Four-dimensional integrable theories},
 hep-th/9410147;
\bibitem{mywork} A. Koubek  {\em Nucl. Phys.} {\bf B435} (1995), 703;
 {\em Phys. Lett.} {\bf B346} (1995), 275;
\bibitem{Karowski}
M.  Karowski, P. Weisz, {\em Nucl. Phys.} {\bf B139} (1978), 445;
\bibitem{nankai}
  F.A. Smirnov, {``Form--Factors in Completely Integrable Models of
    Quantum Field Theory''}, (World Scientific 1992); F. A. Smirnov,
  in {\em Introduction to Quantum Group and Integrable Massive Models
    of Quantum Field Theory}, Nankai Lectures on Mathematical Physics,
  World Scientific 1990;
\bibitem{mywork} A. Koubek  {\em Nucl. Phys.} {\bf B435} (1995), 703;
A. Koubek {\em Phys. Lett.} {\bf B346} (1995), 275;
\bibitem{Zam} A.B.
  Zamolodchikov, in {\em Advanced Studies in Pure Mathematics} {\bf
    19} (1989), 641; {\it Int. J. Mod. Phys.}{\bf A3} (1988), 743;
\bibitem{kedem} R. Kedem, T.R. Klassen, B.M. McCoy and E. Melzer,
{\em Phys. Lett.} {\bf B304} (1993), 263; {\em Phys. Lett.} {\bf B307}
(1993), 68;
\bibitem{spinon} P. Bouwknegt, A.W.W. Ludwig and K. Schoutens,
 {\em Phys.Lett.} {\bf B338} (1994), 448; hepth/9504074;
\end{thebibliography}
\end{document}